\documentclass[iop,revtex4]{emulateapj}

\usepackage{times}
\usepackage{natbib}
\usepackage[backref,breaklinks,colorlinks,citecolor=cyan]{hyperref} 
\usepackage{booktabs}
\usepackage{graphicx}
\usepackage{bm}
\usepackage{multirow}
\usepackage{enumitem}
\usepackage{amsmath}
\usepackage{float}
\usepackage[caption=false]{subfig}

\shorttitle{Dual quasars at close separation with Subaru HSC}
\shortauthors{Silverman et al.}

\begin{document}


\title{Dual supermassive black holes at close separation revealed by the Hyper Suprime-Cam Subaru Strategic Program}


\author{John D. Silverman\altaffilmark{1,2}, Shenli Tang\altaffilmark{1,3}, Khee-Gan Lee\altaffilmark{1}, Tilman Hartwig\altaffilmark{1,3}, Andy Goulding\altaffilmark{4}, Michael A. Strauss\altaffilmark{4}, Malte Schramm\altaffilmark{5}, Xuheng Ding\altaffilmark{6}, Rogemar Riffel\altaffilmark{7,8}, Seiji Fujimoto\altaffilmark{9}, Chiaki Hikage\altaffilmark{1}, Masatoshi Imanishi\altaffilmark{5}, Kazushi Iwasawa\altaffilmark{10}, Knud Jahnke\altaffilmark{14}, Issha Kayo\altaffilmark{12}, Nobunari Kashikawa\altaffilmark{2}, Toshihiro Kawaguchi\altaffilmark{13}, Kotaro Kohno\altaffilmark{2}, Wentao Luo\altaffilmark{1}, Yoshiki Matsuoka\altaffilmark{14}, Yuichi Matsuda\altaffilmark{5}, Tohru Nagao\altaffilmark{14}, Masamune Oguri\altaffilmark{3}, Yoshiaki Ono\altaffilmark{8}, Masafusa Onoue\altaffilmark{11}, Masami Ouchi\altaffilmark{5,8}, Kazuhiro Shimasaku\altaffilmark{2}, Hyewon Suh\altaffilmark{15}, Nao Suzuki\altaffilmark{1}, Yoshiaki Taniguchi\altaffilmark{17}, Yoshiki Toba\altaffilmark{18}, Yoshihiro Ueda\altaffilmark{18}, Naoki Yasuda\altaffilmark{1}}
\altaffiltext{1}{Kavli Institute for the Physics and Mathematics of the Universe, The University of Tokyo, Kashiwa, Japan 277-8583 (Kavli IPMU, WPI)}
\altaffiltext{2}{Department of Astronomy, School of Science, The University of Tokyo, 7-3-1 Hongo, Bunkyo, Tokyo 113-0033, Japan}
\altaffiltext{3}{Institute for Physics of Intelligence, School of Science, The University of Tokyo, Bunkyo, Tokyo 113-0033, Japan}
\altaffiltext{4}{Department of Astrophysical Sciences, Princeton University, 4 Ivy Lane, Princeton, NJ 08544, USA}
\altaffiltext{5}{National Astronomical Observatory of Japan, 2-21-1 Osawa, Mitaka, Tokyo 181-8588, Japan}
\altaffiltext{6}{Department of Physics and Astronomy, University of California, Los Angeles, CA, 90095-1547, USA}
\altaffiltext{7}{Department of Physics \& Astronomy, Johns Hopkins University, Bloomberg Center, 3400 N. Charles St, Baltimore, MD 21218, USA}
\altaffiltext{8}{Universidade Federal de Santa Maria, CCNE, Departamento de F\'isica, 97105-900, Santa Maria, RS, Brazil}
\altaffiltext{9}{Institute for Cosmic Ray Research, The University of Tokyo, 5-1-5 Kashiwanoha, Kashiwa, Chiba 277-8582, Japan}
\altaffiltext{10}{ICREA and Institut de Ci\`encies del Cosmos, Universitat de Barcelona, IEEC-UB, Mart\'i i Franqu\`es, 1, 08028 Barcelona, Spain.}
\altaffiltext{11}{Max-Planck-Institut f\"ur Astronomie, K\"onigstuhl 17, D-69117 Heidelberg, Germany}
\altaffiltext{12}{Department of Liberal Arts, Tokyo University of Technology, Ota-ku, Tokyo 144-8650, Japan}
\altaffiltext{13}{Department of Economics, Management and Information Science, Onomichi City University, Hisayamada 1600-2, Onomichi, Hiroshima 722-8506, Japan}
\altaffiltext{14}{Research Center for Space and Cosmic Evolution, Ehime University, 2-5 Bunkyo-cho, Matsuyama, Ehime 790-8577, Japan}
\altaffiltext{15}{Subaru Telescope, National Astronomical Observatory of Japan (NAOJ), National Institutes of Natural Sciences (NINS), 650 North A'ohoku place, Hilo, HI 96720, USA}
\altaffiltext{16}{The Open University of Japan, 2-11 Wakaba, Mihama-ku, Chiba 261-8586, Japan}
\altaffiltext{17}{Department of Astronomy, Kyoto University, Kitashirakawa-Oiwake-cho, Sakyo-ku, Kyoto 606-8502, Japan}
\email{silverman@ipmu.jp}



\begin{abstract}
The unique combination of superb spatial resolution, wide-area coverage, and deep depth of the optical imaging from the Hyper Suprime-Cam (HSC) Subaru Strategic Program is utilized to search for dual quasar candidates. Using an automated image analysis routine on 34,476 known SDSS quasars, we identify those with two (or more) distinct optical point sources in HSC images covering 796 deg$^2$. We find 421 candidates out to a redshift of 4.5 of which one hundred or so are more likely after filtering out contaminating stars. Angular separations of 0\farcs6--4\farcs0~ correspond to projected separations of 3--30 kpc, a range relatively unexplored for population studies of luminous dual quasars. Using Keck-I/LRIS and Gemini-N/NIFS, we spectroscopically confirm three dual quasar systems at $z<1$, two of which are previously unknown out of eight observed, based on the presence of characteristic broad emission lines in each component, while highlighting that the continuum of one object in one of the pairs is reddened. In all cases, the [OIII]$\lambda$5007 emission lines have mild velocity offsets, thus the joint [OIII] line profile is not double-peaked. We find a dual quasar fraction of $0.26\pm0.18\%$ and no evidence for evolution. A comparison with the Horizon-AGN simulation seems to support the case of no evolution in the dual quasar fraction when broadly matching the quasar selection. These results may indicate a scenario in which the frequency of the simultaneous triggering of luminous quasars is not as sensitive as expected to the cosmic evolution of the merger rate or gas content of galaxies.
\end{abstract}

\section{Introduction}

As structure in the Universe grows hierarchically, galaxies occasionally experience major mergers with one another \citep{SandersMirabel1996}. Such collisions can induce prodigious levels of star formation \citep[e.g.,][]{Fensch2017} and potentially boost the growth of the central supermassive black hole (SMBH) in each galaxy \citep[e.g.,][]{DiMatteo2005, Hopkins2006}. Cosmological hydrodynamic simulations \citep[e.g.,][]{Volonteri2016,Capelo2017,Rosas-Guevara2019} of galaxy evolution predict that in some cases, the two supermassive black holes in a merging galaxy pair will accrete matter at the same time, and thus appear as a luminous dual quasar.

Searches for dual quasars \citep[e.g.,][]{Hewett1998,Hennawi2010} and pairs of less luminous Active Galactic Nuclei \citep[e.g.,][]{Liu2011,Comerford2012,Imanishi2014,Hou2019,Pfeifle2019} have had some success. Recently, \citet{DeRosa2019} provide a thorough review on the searches for binary supermassive black holes on various scales and across different wavebands. Briefly, detailed investigations of double-peaked [OIII]$\lambda4959,5007$ emitters have enabled estimates of the frequency of dual AGNs \citep[e.g.,][]{Comerford2013} and their host galaxy properties \citep[e.g.,][]{Liu2013,Comerford2015}. However, a fair fraction of these lower luminosity candidates are attributed to gas motions in the narrow-line region due to other effects such as outflows \citep[e.g.,][]{Rosario2011,Fu2012,Comerford2012,Mueller-Sanchez2015}. Only a modest number of more luminous dual quasars have been identified at close separation, i.e., during the final stages in the merger at $\lesssim$20 kpc, that have been identified \citep{Komossa2003,Hennawi2006,Shields2012,Inada2012,More2016,Eftekharzadeh2017,Goulding2019}, due to limitations with instrumental spatial resolution. 

Current and future wide-field deep imaging surveys have the potential to significantly improve upon the number of confirmed dual quasars, particularly at higher luminosity and closer separations. For instance, the wide-area coverage, depth, and superb seeing of the Subaru Strategic Program \citep[SSP; ][]{Aihara2018,Aihara2019} with Hyper Suprime-Cam (HSC; \citealt{Miyazaki2018}) can be used to search for rare and key phases of black hole growth, including dual quasars as signposts of major galaxy mergers. Simulations show that quasar accretion is episodic \citep[e.g.,][]{Capelo2017}, thus the supermassive black holes in a major merger will both shine as a quasar only a small fraction of the merger time. The areal coverage of HSC is now sufficient to detect these rare events in significant numbers. Equally important, the median seeing of the HSC imaging is $0\farcs6$ in the i-band, meaning that dual quasars can be identified with projected separations as small as $\sim$3 kpc at $z\sim0.3$. For those at $z<1$, the underlying host galaxies can be robustly detected with HSC including those with faint structures such as tidal debris from an ongoing merger \citep{Goulding2018}.

To exploit the capabilities of HSC on this topic, we are carrying out a program to search for SDSS quasars, falling within the HSC survey footprint, that are dual quasar candidates based on an automated 2D image detection algorithm to identify multiple compact emission on the scale of the HSC point spread function. These objects were initially identified as single SDSS quasars based on optical photometry and spectroscopy, given the lower spatial resolution of the SDSS data.  Dual quasars with separations $>$2$^{\prime\prime}$ are also missed by SDSS, as the survey could not obtain spectra of close pairs of objects due to a limitation on the minimum distance between two optical fibers on a single spectroscopic plate. Even so, subsequent dedicated spectroscopic followup programs of dual quasar candidates have confirmed 47 cases with most at wider separations \citep[$>2\farcs9$;][]{Eftekharzadeh2017} with essentially no overlap with the HSC SSP program.   

Using the Subaru HSC SSP imaging, we report here on the identification of dual quasar candidates having projected separations between 3 and 30 kpc. An optical spectroscopic campaign is underway using Keck-I/LRIS, Subaru/FOCAS and Gemini (GMOS+NIFS) to confirm a statistical sample of dual quasars.  Here, we present our first three spectroscopic confirmations based on Keck and Gemini/NIFS observations. In Section~\ref{sec:selection}, we describe the procedure to identify dual quasar candidates using HSC imaging. The followup spectroscopic campaigns are summarized in Section~\ref{sec:spectroscopy} and three confirmed dual quasars are highlighted in Section~\ref{sec:results}. Based on these initial results, we give a preliminary estimate of the dual fraction of luminous quasars (Section~\ref{sec:dualfrac}).

We discuss our results in the context of cosmological hydrodynamic simulations that span volumes capable of providing expectations on the rate of dual AGN activity and its evolution with redshift \citep{Rosas-Guevara2019,Volonteri2016,Steinborn2016}. In particular, dual quasar fractions from the Horizon-AGN simulation (M. Volonteri - private communication) are presented with parameters (i.e., redshift, quasar luminosity, ratio between the luminosity of the two quasars, projected physical separation) broadly matched to our sample (Section~\ref{sec:dualfrac}). However, any firm conclusions from such comparisons require a larger spectroscopic sample and better understanding of the selection function for both the observational and simulated samples. Throughout this paper, we use a Hubble constant of $H_0 = 70$ km s$^{-1}$ Mpc$^{-1}$ and cosmological density parameters $\Omega_\mathrm{m} = 0.3$ and $\Omega_\Lambda = 0.7$. 

\begin{figure}
\epsscale{1.2}
\plotone{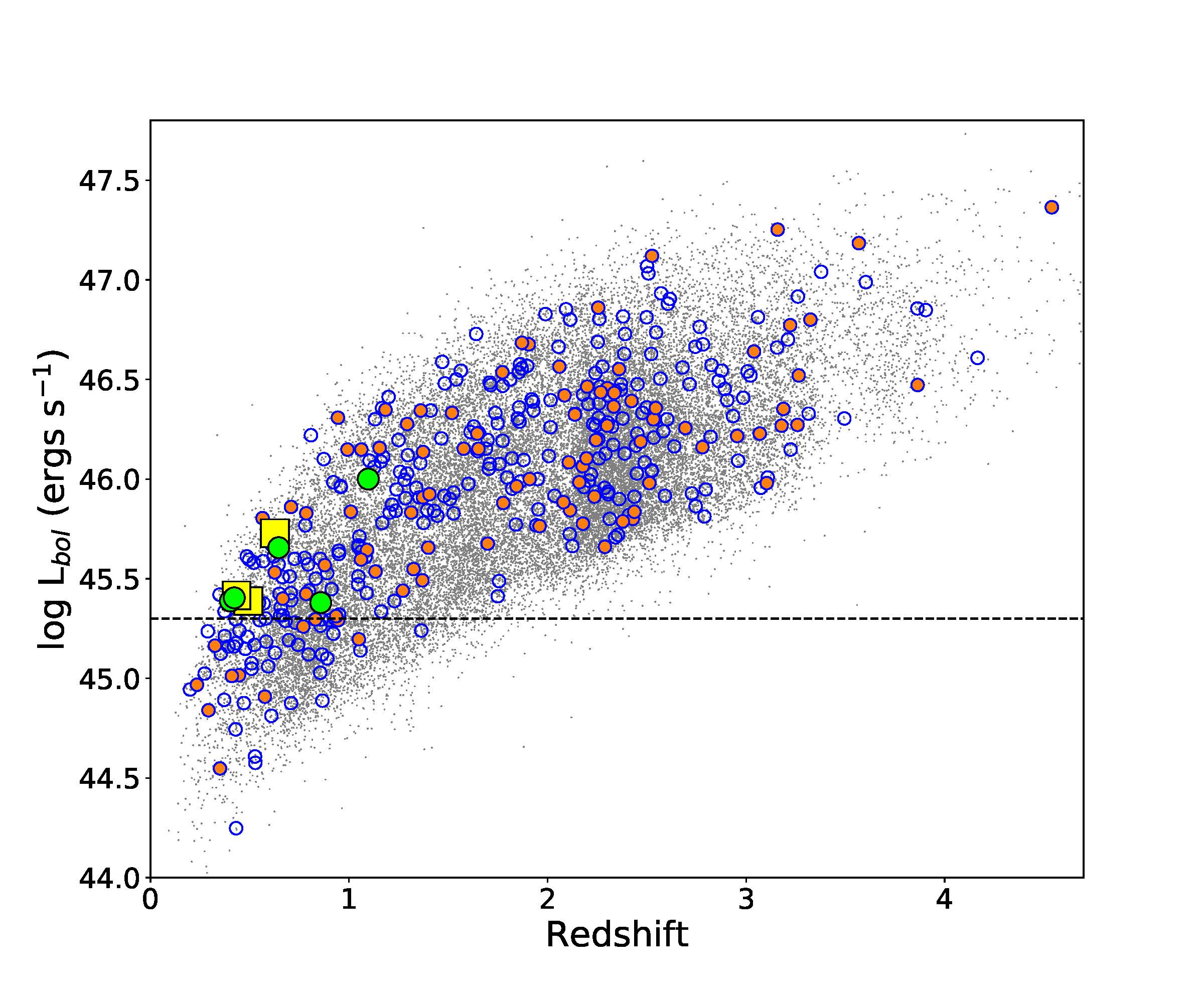}
\caption{SDSS quasar population imaged by Subaru HSC and identified as dual quasar candidates: Distribution of bolometric luminosity ($L_{bol}$), as tabulated in the SDSS DR14 catalog (small grey points), as a function of redshift. The initial 421 dual candidates are marked either as open blue circles (including those having optical colors consistent with stars), and red circles indicate those (116) with the companion having $g-r<1$ that removes many contaminating stars. The yellow squares mark the location of our first three spectroscopically confirmed dual quasars presented in this work. Other spectroscopic targets are shown by the green circles. The horizontal dashed line indicates the luminosity limit of the primary target used to calculate the dual quasar fraction.}
\label{fig:sample}
\end{figure}

\begin{figure*}
\epsscale{0.9}
\plotone{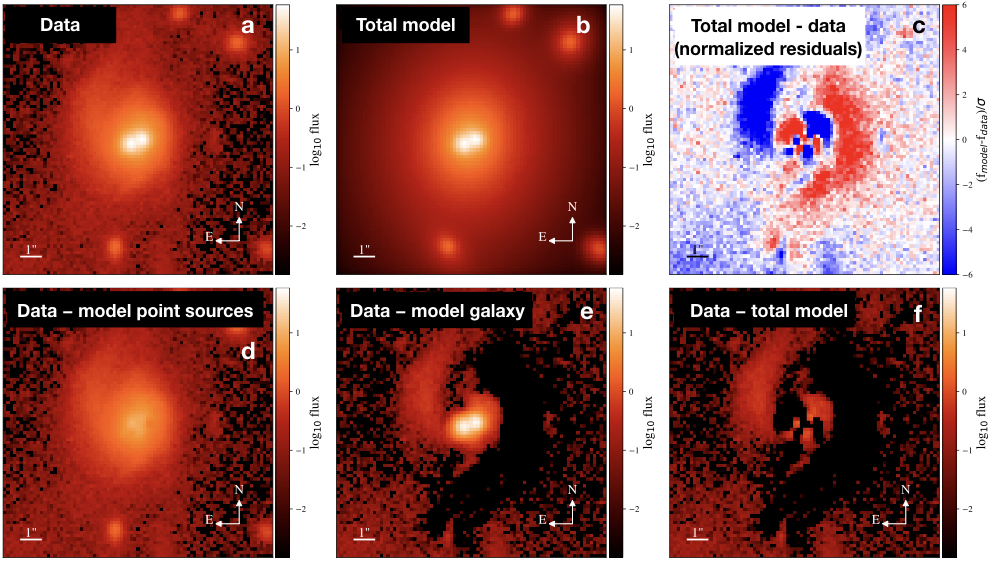}
\caption{Example of our two-dimensional image decomposition of a dual quasar candidate (SDSS J1416+0033) with two point-source components and a S\'{e}rsic model. The panels are as follows: (a) original $i$-band HSC image (logarithmic stretch) with a spatial resolution of $0\farcs51$, (b) smooth model galaxy plus two point-source components convolved with the PSF, (c) residual emission after subtracting the best-fit model from the data and dividing by the error map (linear scaling), (d) data minus the two point sources (image of the host galaxy only), (e) data minus the smooth host model (two quasars only) and (f) data minus the model including all components (residual emission) similar to panel c. A scale bar of 1$^{\prime\prime}$ is shown in each panel.}
\label{fig:decomp_example}
\end{figure*}

\section{Identifying dual quasar candidates from the luminous SDSS quasar population with Subaru HSC}
\label{sec:selection}

The imaging used in this study is drawn from the second public data release of the HSC SSP program \citep{Aihara2019}, including 796 deg$^2$ of $i$-band imaging. We include all imaging with at least one 200 second exposure, so not all data reach the final depth of 26.2 mag (5$\sigma$; AB). Our initial selection of dual quasar candidates is based on the $i$-band due to the quality of the seeing in that band. We then use the images taken in the other four broad-band filters \citep[i.e., $g$, $r$, $z$, $y$;][]{Kawanomoto2018} to provide color information. The imaging data are processed through the standard pipeline \citep{Bosch2018} hscPipe Version 6.7.

The SDSS DR14 v.4.4 quasar catalog \citep{Paris2018} contains 526,357 spectroscopically confirmed quasars with spectroscopic redshift up to $z\sim5$ (Figure~\ref{fig:sample}). The SDSS quasar selection includes magnitude- and color-selected samples \citep{Richards2002,Ross2012,Myers2015} and objects identified through other means, e.g., radio, X-ray, infrared. Of these, 34,476 quasars have catalog entries in the HSC database within the usable HSC exposure area which are not flagged as saturated ($i_{AB}\gtrsim18$), having a bad pixel, or unable to determine a magnitude based on a model fit. HSC image cutouts of size $60^{\prime\prime}\times60^{\prime\prime}$ are generated for each object, together with the variance image and model PSF.

An automated algorithm in Python is run on each Subaru/HSC i-band cutout image to detect those with multiple optical components. These analysis tools, provided by the open-source package Lenstronomy \citep{Birrer2018}, have been slightly modified to process HSC images from versions used to de-blend AGN and host galaxy emission based on HST imaging \citep{Ding2020}. We initially detect peaks in each image based on emission above a given signal-to-noise ratio and over a number of contiguous pixels. We then perform a forward modeling of the two-dimensional distribution of the emission. This model includes unresolved emission characteristic of the point-spread function (PSF; \citealt{Coulton2018}) of each quasar, and 2-dimensional S\'ersic profiles for the host galaxy, usually detected for only those quasars with $z < 1$. Accurate models of the PSF for each quasar are of primary importance for this analysis. We utilize the empirical model PFS used for the lensing measurements that have been constructed from stars that fall on each CCD of the main target. \citet{Carlsten2018} presents a detailed analysis of the wavelength dependence of the PSF models. In most cases, the HSC pixel scale of $0\farcs168$ is sufficient to sample the PSF. We simultaneously fit all galaxies in the field-of-view of the image cutout since their extended profiles can contribute flux to the main targets of interest. We also allow for a local constant sky level to account for errors in the global sky subtraction routine of the standard pipeline analysis. In Figure~\ref{fig:decomp_example}, we demonstrate the results for an example dual quasar candidate. This object has been selected to demonstrate our analysis routine and may be an exceptional case as described further below. For each component, our procedure returns the centroid, total flux, S\'ersic parameters (half-light radius, S\'ersic index) and ellipticity. Further details and examples of the fitting procedure of AGNs and their host galaxies using these tools with HSC data will be presented in a paper in preparation. In particular, while the deep limiting sensitivity of the HSC imaging can allow us to tie the growth of the individual SMBHs to the evolutionary state of the galaxy merger for cases at $z<1$, we reserve such analysis to a more focused effort, primarily since the two-dimensional modeling of the host optical emission from merging galaxies may require additional components in the fitting routine due to complexity in their light distribution.

Dual quasar candidates are those with multiple optical components with a separation between $0\farcs6$ and 4$^{\prime\prime}$. The lower bound is determined by the spatial resolution of the imaging while the upper bound is arbitrarily set to reach separations of a few tens of kpc while minimizing the level of contamination. The procedure initially identified 452 candidates at these separations (Table~\ref{tab:sample}). Visual inspection caused us to reject 27 objects as artifacts. In addition, four objects were found to be gravitational lenses, including one identified through our Keck spectroscopy (see below). Thus, there are 421 dual quasar candidates with 385 having a flux ratio of 10:1 or less, prior to consideration of the contamination by foreground stars (Section~\ref{sec:stars}). In Figure~\ref{fig:sample} we show the bolometric luminosities of this sample relative to the overall parent quasar population as a function of redshift. In Table 1, we provide the sample size at each step in the selection process.

\begin{figure}
\epsscale{1.1}
\plotone{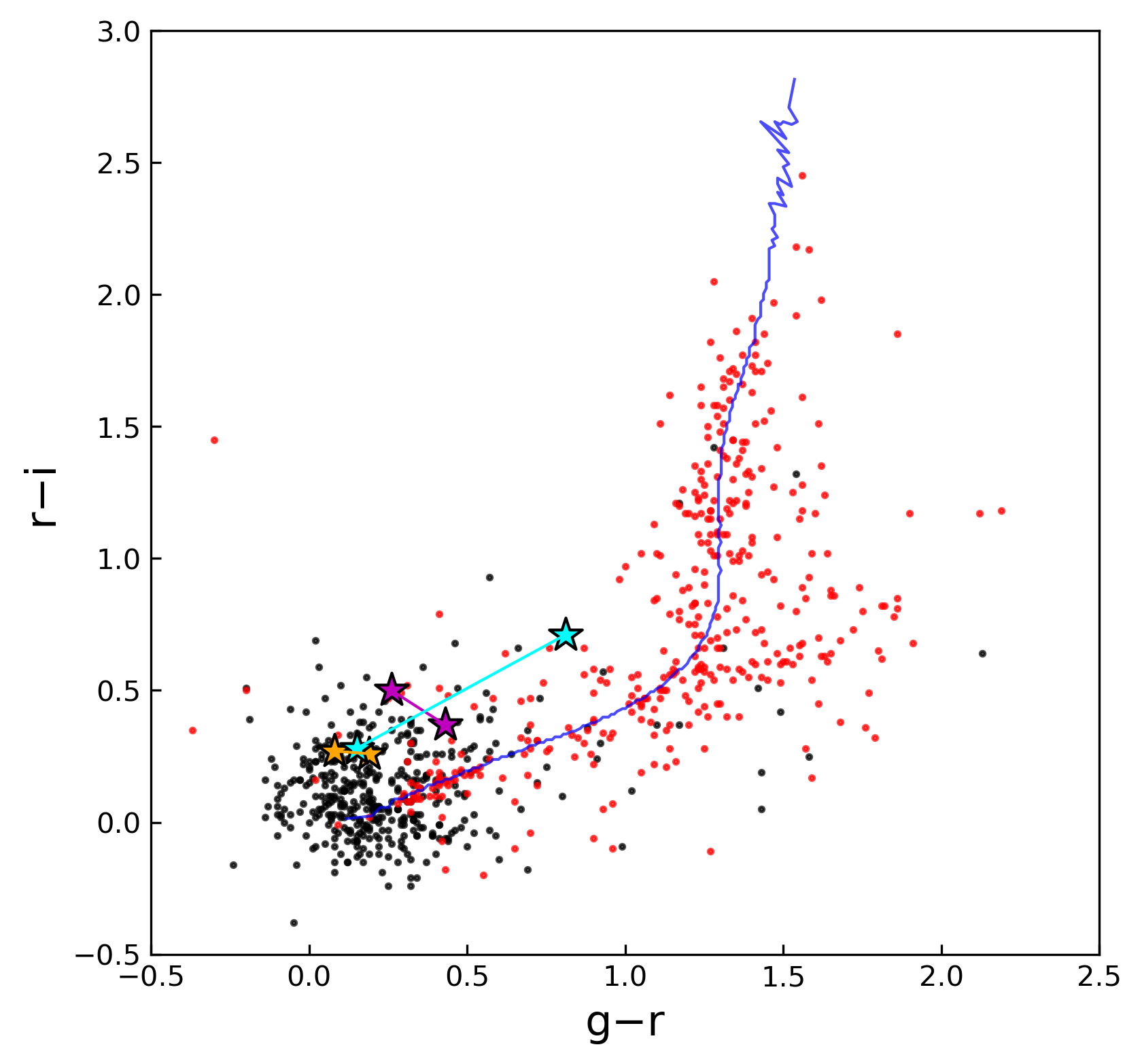}
\caption{Optical colors ($g-r$ vs. $r-i$) of the individual sources (primary SDSS target: small black circles; companion: small red circles) within our dual quasar sample. The stellar locus \citep{Covey2007} is shown by the blue line, adjusted slightly to match the HSC photometric system. The three spectroscopically-confirmed dual quasars are shown as larger stars with each pair connected by a solid line. A larger open circle marks the companions in our sample having spectroscopy.}
\label{fig:colors}
\end{figure}

\begin{deluxetable}{lr}
\tabletypesize{\scriptsize}
\tablecaption{Sample selection\label{tab:sample}}
\tablehead{\colhead{Category}&\colhead{Number of objects}}
\startdata
SDSS DR14 quasar catalog&526357\\
Imaged by the HSC wide-area survey&34476\\
Dual quasar candidates with 0.6--4$^{\prime\prime}$ separation&452\\
"~~~~~~~~~~~~~~~~~~~~~~~~~~~~~~~~~~~" after visual inspection&425\\
"~~~~~~~~~~~~~~~~~~~~~~~~~~~~~~~~~~~" minus known lenses&421\\
"~~~~~~~~~~~~~~~~" with 5-band photometry available&401\\  
"~~~~~~~~~~~~~~~~~~~~~~~~~" having flux ratio within 10:1&385\\
"~~~~~" with the companion having $g-r<1.0$&116
\enddata
\end{deluxetable}

\subsection{Stellar contamination}
\label{sec:stars}

We inspect the optical colors of the individual components of our dual quasar sample (Figure~\ref{fig:colors}) for those (401) having photometry in all five bands. Many of these objects lie close to the stellar locus \citep{Covey2007}, suggesting that foreground stars are a major contaminant. In an attempt to remove some of the contaminating stars, there are 116 dual candidates in which the companion to the target quasar has $g - r < 1$. However, this color cut is not definitive; massive stars can be this blue, and dust-reddened quasars can be redder thus followup spectroscopy is needed to confirm dual quasars as candidates. In Section~\ref{sec:dualfrac}, we implement this simple color selection in a preliminary assessment of the dual quasar fraction.  In future work, we will explore more rigorous methods to remove stars, such as by using the distance to the stellar locus.

\section{Spectroscopic followup}
\label{sec:spectroscopy}

The presence of a pair of close point sources does not prove that a given object is a dual quasar. It could be a single quasar gravitationally lensed by a foreground galaxy, or the chance projection of a star, a compact galaxy, or a quasar at a different redshift. An object can be confirmed as a dual quasar with spatially resolved spectroscopy of the two objects, which requires seeing good enough to separate the two components. Therefore, we have acquired spectroscopic observations of dual quasar candidates in 2019 using Keck-I, Subaru and Gemini. Here, we describe the initial Keck-I/LRIS and Gemini-NIFS spectroscopic observations used to confirm our first dual quasar candidates (Figure~\ref{fig:sample}). A subsequent study (Tang et al. in preparation) will present in detail the spectroscopic programs with Subaru/FOCAS and Gemini GMOS-N.

\begin{figure}
\epsscale{1.1}
\plotone{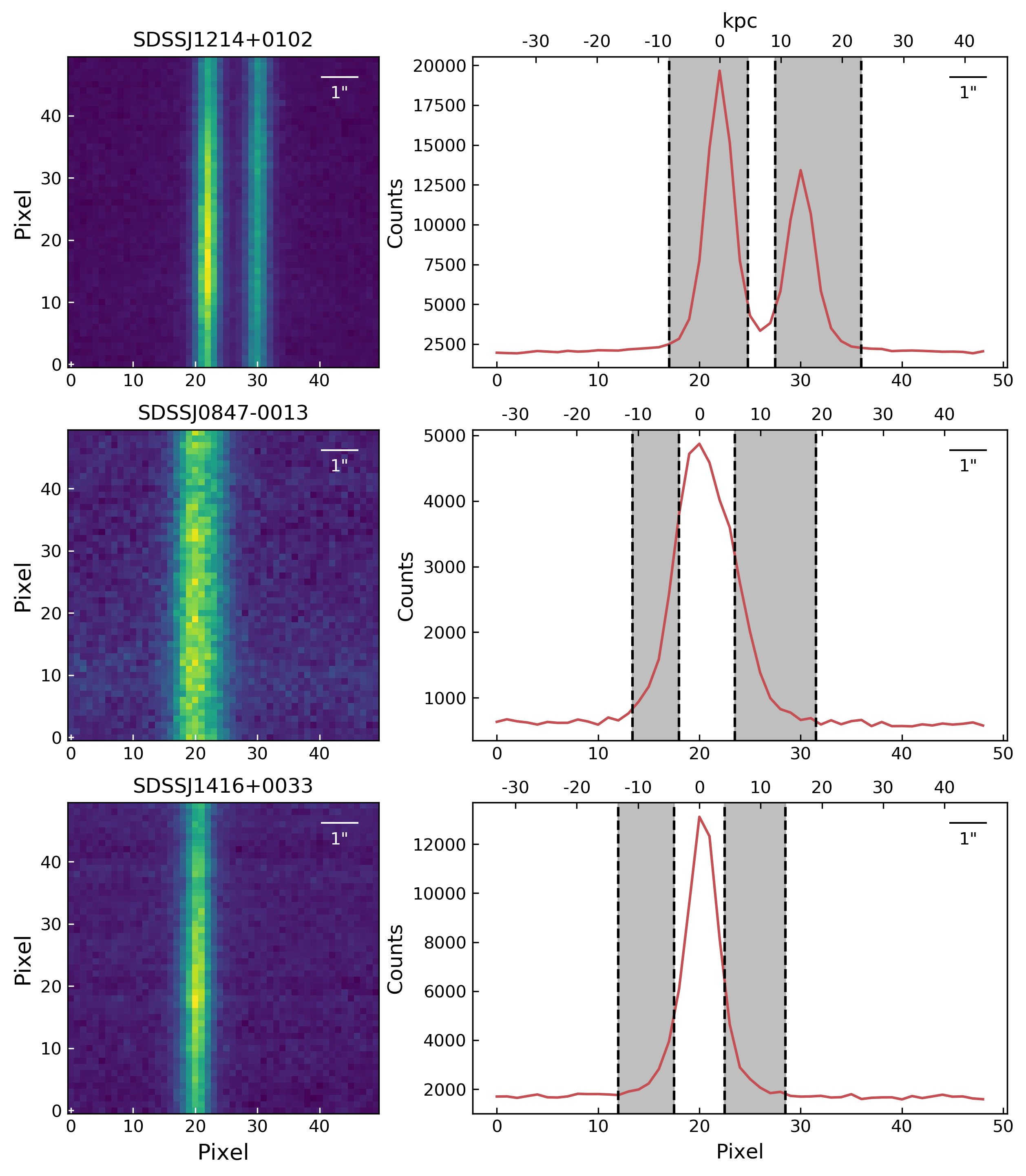}
\caption{Spatial extraction windows for Keck spectroscopy of the three confirmed dual quasar pairs ($top$ SDSSJ1214+0102; $middle$: SDSSJ0847-0013, $bottom$: SDSSJ1416+0033). Images on the left are 2D cutouts surrounding the Mg II emission line. The x-axis is the spatial dimension and the y-axis is the spectral dimension. These images are then collapsed along the spectral dimension to illustrate their spatial profiles (right panels) where the grey shaded regions indicate the spatial range over which each quasar spectrum was extracted. While the two components of SDSSJ1214+0102 and SDSS1416+0033 are blended, the wings of the profile avoid contamination between the two components, allowing us to classify each as a different quasar.}
\label{fig:spatial_profiles}
\end{figure}

\begin{figure}
\epsscale{0.95}
\hskip -1.0cm
\plotone{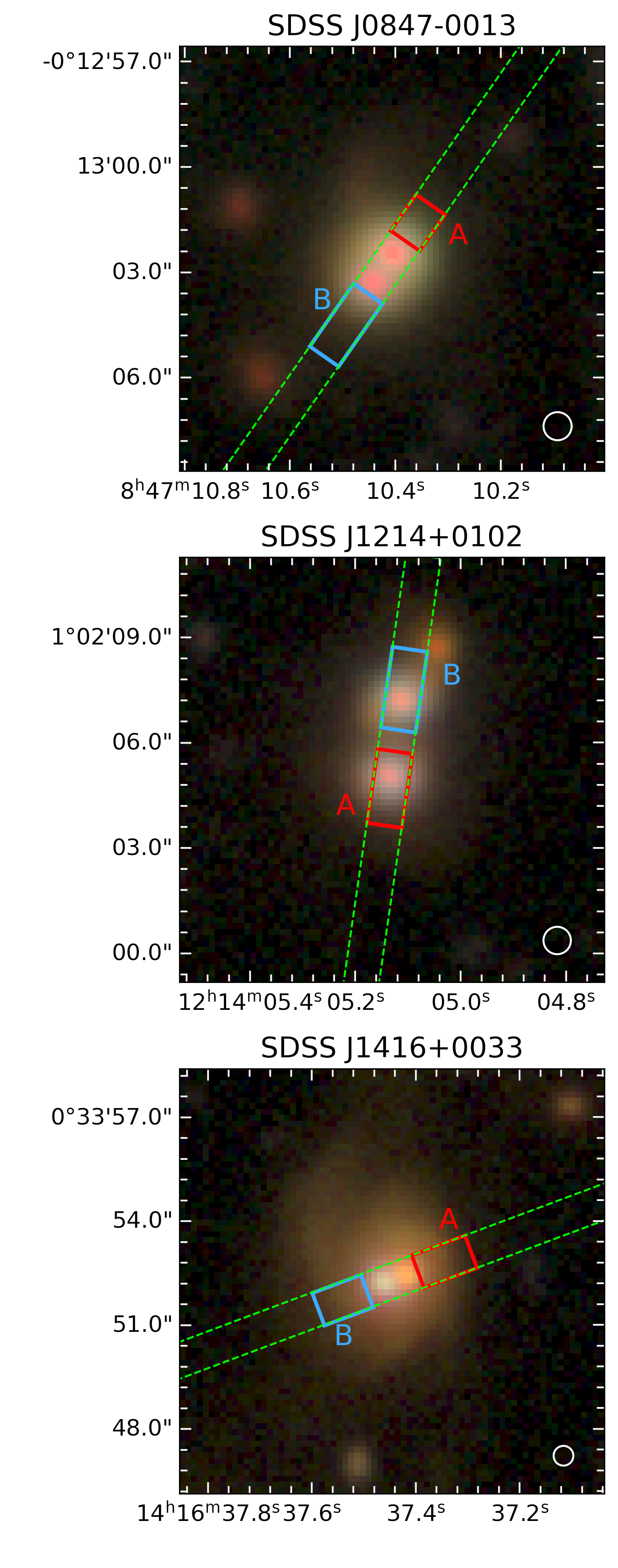}
\caption{Placement of the Keck-I/LRIS slits as shown by the green rectangular regions with the red and blue boxes indicating the regions from which the spectra were extracted. For SDSSJ0847-0013 and SDSS J1416+0033, spectra are extracted from regions (A and B) offset from the cores to minimize blending since the spatial resolution ($\sim0\farcs5$) achieved with Keck was not adequate to fully isolate the two nuclei separated by $0\farcs67$. The circles in the lower right corner of each panel indicate the size (i.e., FWHM) of the PSF for the lowest resolution of the HSC images used to construct the color image.}
\label{fig:slit_placement}
\end{figure}

\begin{figure*}
\epsscale{0.75}
\plotone{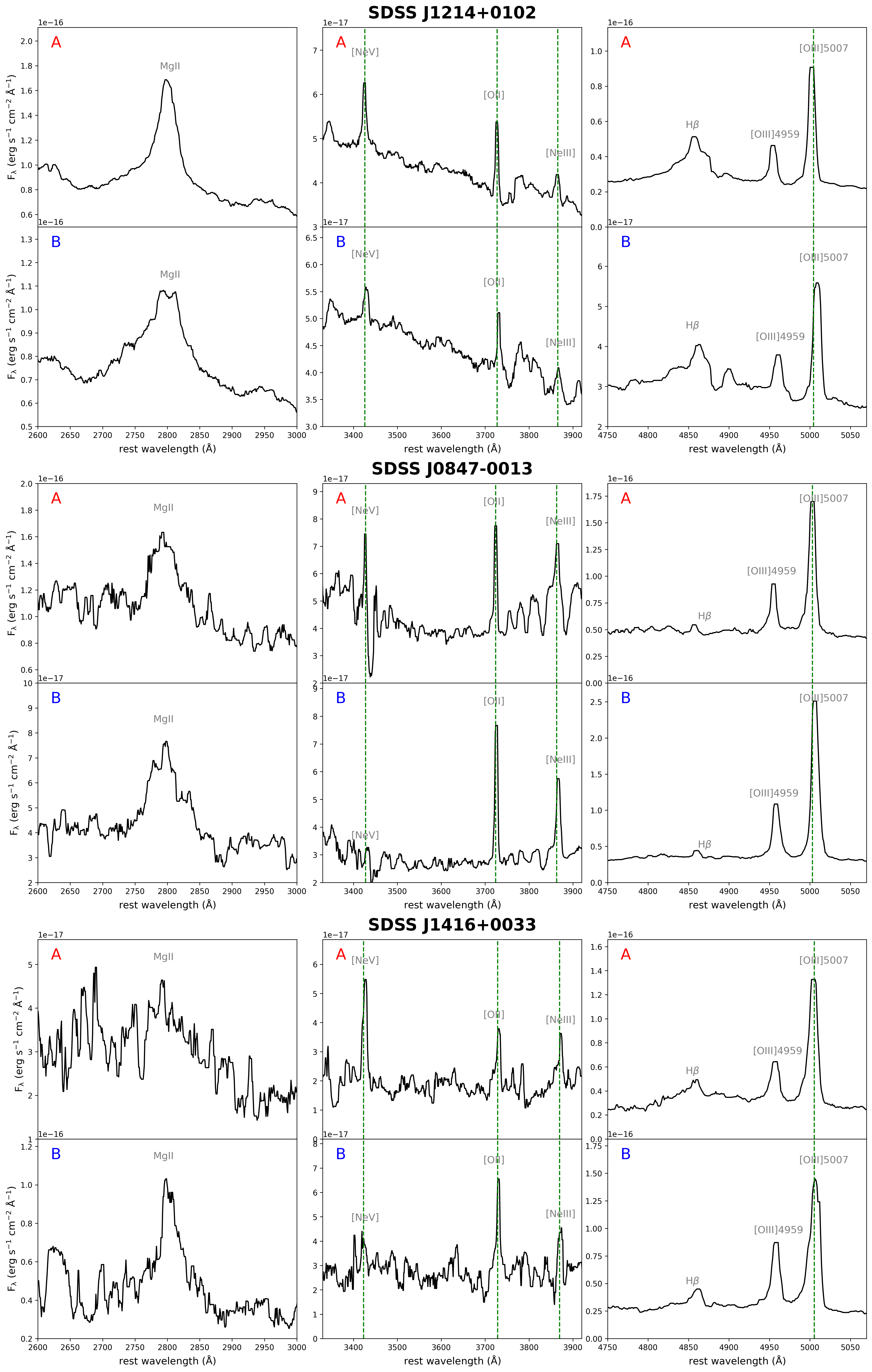}
\caption{Optical spectroscopic confirmation of three dual quasar systems with Keck-I/LRIS. (Top) SDSS J1214+0102, a dual broad-line (type 1--1) quasar at \hbox{$z = 0.4927$} based on the detection of Mg II and H$\beta$ in each component. (Middle) SDSS J0847-0013, a second dual broad-line (type 1--1) quasar (\hbox{$z = 0.6269$}) based on the detection of Mg II and H$\beta$ in each component. (Bottom) SDSS J1416+0033, a dual quasar system at \hbox{z = 0.4336} having an unobscured quasar associated with component B based on the detection of broad Mg II and component A is a quasar as indicated by the strong [NeV] emission. The green vertical line indicates the centroid of [OIII]$\lambda$5007 (rest-frame wavelength) for one of the components of each dual quasar system, highlighting the velocity offsets of the narrow emission lines.
}
\label{fig:spectra}
\end{figure*}

\subsection{Keck-I/LRIS}

We observed eight candidates under suitable conditions with the Low Resolution Imaging Spectrometer (LRIS; \citealt{Oke1995}) on the Keck-I telescope on January 10, 2019. These candidates were selected to be at $z \lesssim 1$, have flux ratios of 3:1 or smaller, and be observable at either the beginning or end of the nights to not conflict with targets for the main program. The flux ratio cut was chosen so that both components would be bright enough to acquire spectra with sufficient signal-to-noise ratio in a 10--15 minute exposure. The 600/7500 grism was used, giving spectral coverage of 3200--5500 \AA~(blue channel) and 5500--8500 \AA~(red channel). Standard stars were observed for flux calibration.

The spectra were taken under a range of seeing conditions ($0.4 - 1\farcs0$). In five cases, the two components could be cleanly separated while in others significant blending occurred (see Figures~\ref{fig:spatial_profiles} and~\ref{fig:slit_placement}). While we were not able to separate the emission into two components for all candidates, we could extract spectra from the wings of the spatial profile in some cases, allowing us to classify each component.

Out of the eight candidates observed with Keck-I, three (SDSSJ0847-0013, SDSSJ1214+0102 and SDSSJ1416+0033) are confirmed as pairs of broad-line quasars (Figure~\ref{fig:spectra}). SDSSJ0847-0013 was previously identified \citep{Inada2008} while the other two are newly discovered dual quasar systems. The other candidates include one gravitationally-lensed quasar at $z = 1.097$ (Jaelani et al. in preparation), two quasar--star pairs, one triple quasar--star--galaxy system, and one unclassified object due to low signal-to-noise. 






\subsection{Gemini-NIFS}

\begin{figure}
\epsscale{0.75}
\plotone{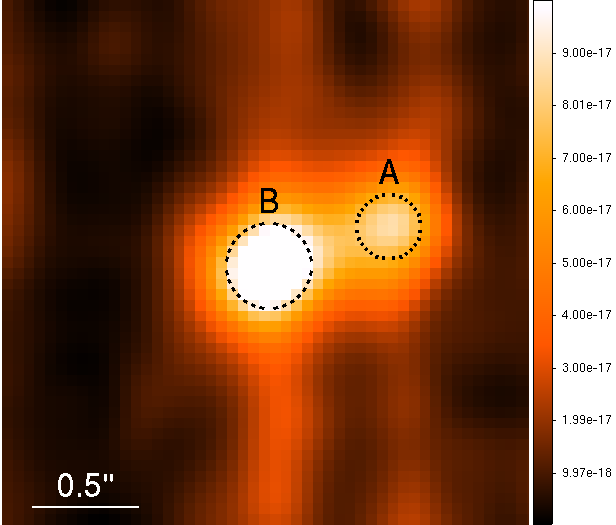}
\epsscale{0.85}
\plotone{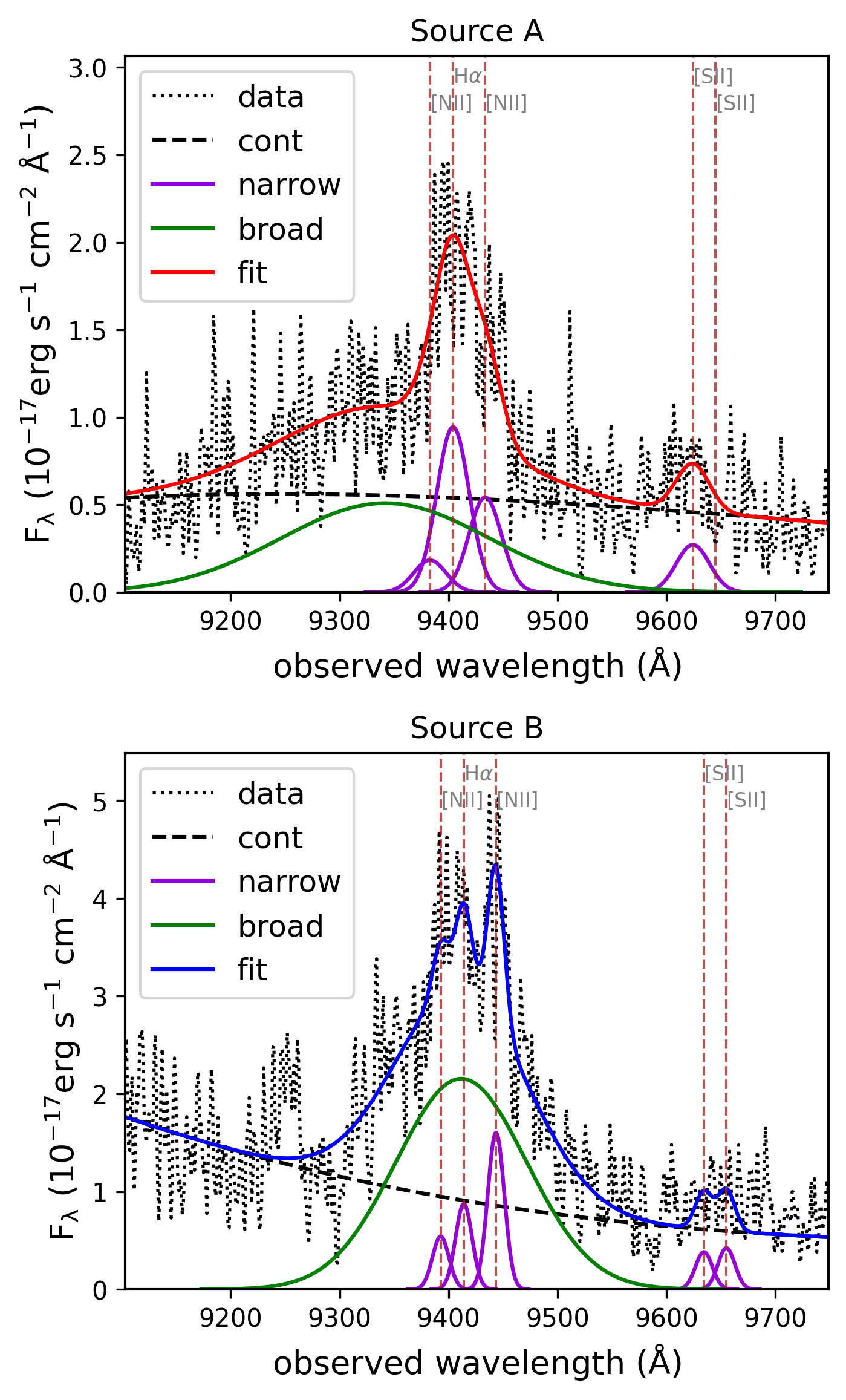}
\caption{Gemini/NIFS $z$-band IFU observations of the central region of SDSS J1416+0033. An image of a spectral region containing H$\alpha$+[NII] is shown on top. North is up and East is to the left. The vertical stripes are likely due to scattered light along the NIFS slices that do not impact the results of this work. The spectrum of each nucleus displayed in the middle and bottom panels with arbitrary flux units of ergs s$^{-1}$ cm$^{-2}$ \AA$^{-1}$. The spectral features have been modeled with Gaussian components as shown in color. The continuum fit is displayed by a dashed black line. The sum of all model components is given in blue.}
\label{fig:nifs}
\end{figure}

SDSS J1416+0033 (Fig.~\ref{fig:decomp_example}), whose Keck spectrum shown in Figure~\ref{fig:spectra} was compelling but not definitive evidence of being a dual quasar, was observed during Director's Discretionary Time (GN-2019A-DD-102) on April 24, 2019 with Gemini North's Near-Infrared Integral Field Spectrometer (NIFS; \citealt{McGregor2003}). In seeing-limited mode, we targeted the H$\alpha$+[NII] spectral region in the z-band to identify emission from the broad line region of each nuclei separately (Figure~\ref{fig:nifs}). The on-source exposure time was 80 min. The wavelength solution was based on an Ar/Xe arc lamp exposure. A dome flat exposure was used to flat-field the image while a single standard star observation allowed us to remove telluric absorption and flux calibrate the spectra. The NIFS data reduction was accomplished following the standard procedure, including the trimming of the images, flat-fielding, cosmic ray rejection, sky subtraction, wavelength and s-distortion calibrations, telluric absorption cancellation, and flux calibration. The construction of the data cubes for individual exposures was done using an angular sampling of 0\farcs05$\times$0\farcs05, and mosaicing of the individual data cubes using as a reference the position of the brightest nucleus -- for details see \citet{Riffel2008}. Following the procedures described in \citet{Menezes2014}, we resampled the final data cube to 0\farcs025 width spaxels, applied a Butterworth spatial band-pass filter to remove high-frequency noise, executed a Richardson-Lucy deconvolution using the flux distributions of the telluric standard star as the point-spread function 
and finally we resampled the data cube back to the 0\farcs05 width spaxels.

\begin{deluxetable}{llll}
\tabletypesize{\scriptsize}
\tablecaption{Spectroscopically-confirmed dual quasars\label{tab:specsample}}
\tablehead{\colhead{ID}&\colhead{RA}&\colhead{DEC}&\colhead{Redshift}\\
&\colhead{(J2000)}&\colhead{(J2000)}}
\startdata
SDSS~J121405.12+010205.1&12:14:05.13& +01:02:05.0&0.4927\\
SDSS~J084710.40-001302.6&08:47:10.41&-00:13:2.5&0.6269\\
SDSS~J141637.44+003352.2 &14:16:37.46& +00:33:52.2 &0.4336
\enddata
\end{deluxetable}

\section{Results}
\label{sec:results}

We confirm three candidates as dual quasars at $z < 1$ with spectroscopic observations using Keck-I/LRIS and Gemini-N/NIFS (Table~\ref{tab:specsample}). In Figures~\ref{fig:decomp_example},~\ref{fig:1214decomp} and~\ref{fig:0847decomp}, we show the HSC images of the three dual quasars. HSC spatially separates the individual components cleanly in each case. After subtracting the point source components of the model, the host galaxies show tidal features and asymmetries indicative of a merger. These dual quasar systems increase the numbers of known galaxy mergers in which the individual nuclei are at close projected separations (4 -- 24 kpc) and both SMBHs are accreting at quasar-like luminosities: the primary target has $L_{bol}>10^{45}$ ergs s$^{-1}$ (Figure~\ref{fig:sample}) and the companion more than 0.1 as luminous as the primary. At these separations, most studies have reported dual AGNs at lower luminosities ($L_{bol} \lesssim 10^{44}$ ergs s$^{-1}$; \citealt{Liu2013,Comerford2015,Hou2019,Pfeifle2019}) with most well below this limit. Here, we describe these three examples while highlighting some common features (i.e., relatively low [OIII]$\lambda$5007 velocity offsets). We further derive the stellar masses of the host galaxies from de-blended HSC photometry using CIGALE \citep{Boquien2019}, a delayed star-formation history, and a Chabrier Initial Mass Function \citep{Chabrier2003}. While the individual black hole mass estimates will be presented elsewhere (Tang et al. in preparation), we do mention the range of black hole mass spanned by the sample in Section~\ref{sec:dualfrac}. We order the three examples by decreasing projected physical separation, possibly representing different stages in a merger sequence of galaxies and their SMBHs.


\subsection{SDSS J1214+0102}

\begin{figure}
\epsscale{1.2}
\plotone{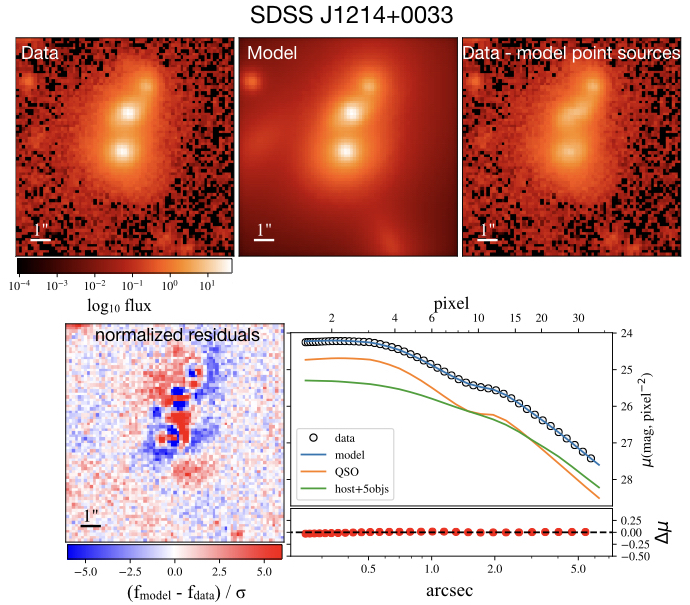}
\caption{SDSS J1214+0033: 2D decomposition of the HSC $i$-band image with panels as described in Figure~\ref{fig:decomp_example}. The spatial resolution is $0\farcs52$. In addition, we show the surface brightness profile and the contribution of the two quasars, host galaxy, and neighboring galaxies.}
\label{fig:1214decomp}
\end{figure}

This system is a newly discovered dual quasar at \hbox{$z = 0.4927$} with a projected separation of $2\farcs2$, corresponding to a distance of 13.2 kpc. As shown in Figure~\ref{fig:1214decomp} (top right panel), there is strong evidence for an underlying massive host galaxy (Table~\ref{tab:decomp}) associated with each quasar and an additional companion to the northwest. Thus this system is likely in the process of merging. While the two main components of SDSSJ 1214+0102 each have broad H$\beta$ and MgII emission (Figure~\ref{fig:spectra}; top row), the line widths are different (MgII full-width-half-maximum of 4400 (A) and 7000 (B) km s$^{-1}$). Furthermore, there is a velocity offset of 442 km s$^{-1}$ between the two components in the [OIII]$\lambda$5007 emission line, which traces the kinematics of the warm ionized gas on the larger scale of the host galaxy. Thus, this system is not two images of the same quasar in a gravitational lens. This velocity offset is too small to be identified as two separate systems in the SDSS joint spectrum of this object, and was thus not found in searches of quasars with double-peaked [OIII] lines \citep{Liu2011,Comerford2012}. We conclude that the three galaxy components, with stellar mass ratio of 2.6:1.3:1 (Table~\ref{tab:decomp}) are undergoing a major merger.  The two most massive galaxies have SMBHs shining as quasars.  The system is at an intermediate stage, with the three galaxies distinguishable, but will presumably merge to a single galaxy in much less than a Hubble time.


\subsection{SDSS J0847-0013}

\begin{figure}
\epsscale{1.18}
\plotone{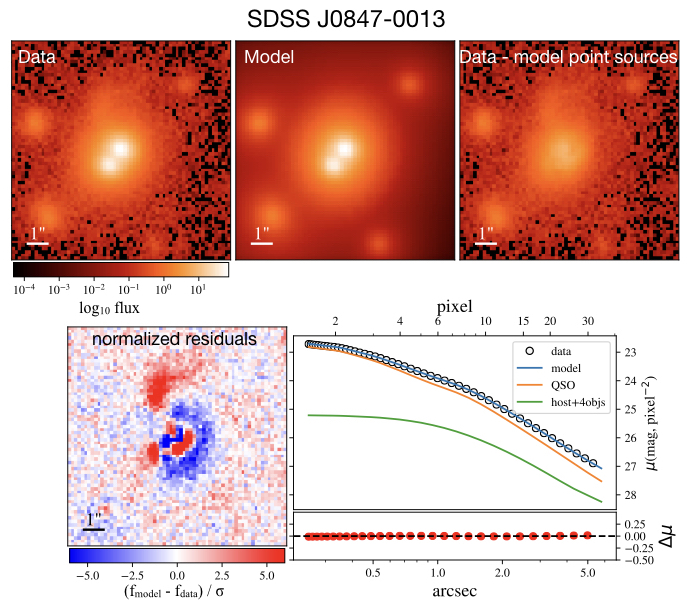}
\caption{SDSS J0847-0013: 2D decomposition of the HSC $i$-band image with panels as described in Figures~\ref{fig:decomp_example} and~\ref{fig:1214decomp}. The spatial resolution is $0\farcs61$}
\label{fig:0847decomp}
\end{figure}

HSC imaging (Fig.~\ref{fig:0847decomp}) shows a single galaxy (right panel) with two bright nuclei separated by 1$^{\prime\prime}$ (left panel), thus a projected separation of 6.8 kpc. The single host galaxy has a stellar mass of $8.5\pm4.6\times10^{10}$ M$_{\odot}$ (Table~\ref{tab:decomp}) based on a S\'{e}rsic model fit. In addition, there is diffuse emission on scales of a few arcseconds, faint tidal features to the north and four nearby faint companion galaxies; this system is clearly undergoing a merger. In fact, SDSS J0847-0013 was previously confirmed as a dual quasar in a search for lensed quasars \citep{Inada2008}, although no host galaxy was detected in the original discovery images. The published spectra show broad MgII emission lines in each component with different line profiles, thus ruling out the possibility that this system is a lensed quasar. Our Keck-I/LRIS spectroscopy (Fig.~\ref{fig:spectra}; middle panels) confirms the presence of the two dissimilar broad MgII profiles. The Keck spectra include H$\beta$ and [OIII]$\lambda$5007 as well. Both H$\beta$ lines are broad and are similar in width to the MgII lines. The [OIII] lines differ in velocity along the line of sight by 194 km s$^{-1}$, too small to show a double-peaked [OIII] profile, as in SDSS J171544.05+600835.7 \citep{Comerford2011}. This is a common characteristic of all three dual quasars in the sample presented here. 

\subsection{SDSS J1416+0033}

As shown in Figure~\ref{fig:decomp_example}, this galaxy at $z = 0.4336$ has undergone a major merger as indicated by tidal structures and a double nucleus. The HSC color image (Fig.~\ref{fig:spectra}, bottom panel) shows two nuclei, one blue and the other red (Fig.~\ref{fig:colors}), separated by only $0\farcs67$, corresponding to a distance of 3.9 kpc. Based on a S\'{e}rsic model fit, the single host galaxy has a stellar mass of $1.6\pm5.3\times10^{11}$ M$_{\odot}$ (Table~\ref{tab:decomp}). At larger separations, there are nearby galaxies with photometric redshifts consistent with J1416+0033 being in a galaxy group.  

Since the optical emission from the two nuclei seen in the Keck spectra (Fig.~\ref{fig:spatial_profiles}, bottom panels) is significantly blended, the regions from which the spectra were extracted were chosen to be offset from each other (Fig.~\ref{fig:spectra}, bottom panel) to minimize contaminating light from each nucleus. This method is effective due to the acceptable seeing conditions ($\sim0\farcs5$), brightness of the two quasars, and the relatively high signal-to-noise spectra in the outer wings of the spatial profiles. 

The Keck spectra reveal a broad MgII and H$\beta$ (Fig.~\ref{fig:spectra}) emission lines associated with the primary blue nucleus (component B). The red nucleus (component A) has weaker evidence from the Keck spectra of a broad emission line in MgII and H$\beta$, but shows prominent narrow lines of high-ionization species, [OIII]$\lambda4959, 5007$ and [NeV], likely indicative of photoionization from a quasar, particularly the presence of strong [NeV] emission \citep{Gilli2010,Mignoli2013,Yuan2016}. Thus, this component is moderately obscured by dust; the [NeV] arises from the narrow-line region of a quasar, photo-ionized by the accretion disk around the central SMBH. Here, the [OIII]$\lambda$5007 lines in the two components are separated by 113 km s$^{-1}$, again not separated large enough to be split into two distinct peaks.

\begin{figure}
\epsscale{0.9}
\plotone{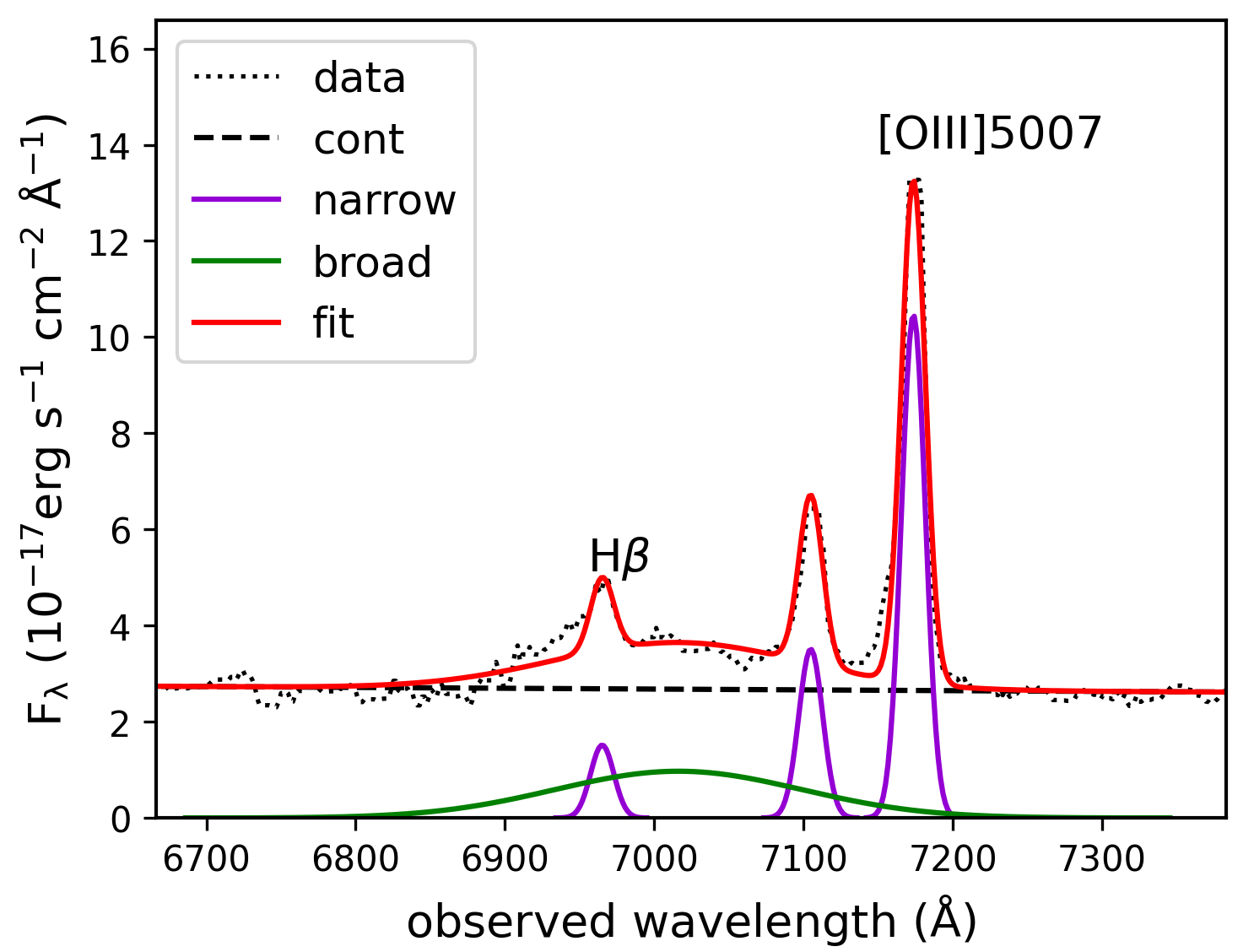}
\caption{Model Gaussian fits to the [OIII]$\lambda4959, 5007$ and H$\beta$ emission lines in component A of SDSS  J1416+0033. The model components are the same as in Figure~\ref{fig:nifs}.}
\label{fig:optfit}
\end{figure}

The coverage of the H$\alpha$+[NII] spectral region with Gemini/NIFS appears to confirm the nature of the red nucleus (component A shown in the bottom panel of Figure~\ref{fig:slit_placement}). In Figure~\ref{fig:nifs}, we show the IFU observation with a spectral extraction of the two nuclei that are now cleanly separated spatially with an effective resolution of $0\farcs28$. In the middle panel, the spectrum of the fainter nucleus (component A) to the Northwest has a clear detection of H$\alpha$ and [NII]. From Gaussian fits to the emission-line complex, the ratios of the narrow components of [NII]$\lambda6584$ to H$\alpha$, and [OIII]$\lambda5007$ to H$\beta$ (Figure~\ref{fig:optfit}), suggest a hard photo-ionizing source. More compellingly, there is evidence for a broad component that is shifted bluewards by 1608 km s$^{-1}$ from the narrower H$\alpha$ component. Based on the broad lines in each component (MgII, H$\beta$, and H$\alpha$) and the narrow emission-line ratios, [NII]$\lambda6584$/H$\alpha$ (Fig.~\ref{fig:nifs}) and [OIII]$\lambda5007$/H$\beta$ (Figure~\ref{fig:optfit}), including the brighter component B (Fig.~\ref{fig:spectra}, bottom panel), we conclude that this system is a dual quasar, in which one component is unobscured while the other one has moderate obscuration but not enough to completely hide the broad-line region. Thus, we classify the pair as a type 1.5, given the hint of a broad component to the weak H$\beta$ line shown in Figure~\ref{fig:spectra}. We note that the velocity offsets between broad emission lines (MgII, H$\beta$ and H$\alpha$) are not yet understood and will be further investigated.
 
We cannot comment yet on how common pairs of obscured and unobscured quasars may be in the dual quasar population based on this one example. Such pairs are likely to be prevalent within the dual quasar population given reports of similar cases confirmed through X-ray observations \citep{Ellison2017,DeRosa2018,Hou2020}, and studies of AGN obscuration in galaxy mergers \cite[e.g.,][]{Ricci2017}. In our case, while some of our dual candidates include a red component, many may be chance projections with foreground stars (Fig.~\ref{fig:colors}). However, those with extended optical emission and/or anomalous colors likely require X-ray followup or further long-slit optical or near-infrared spectroscopy to secure their nature as dual quasars.

\begin{deluxetable*}{llllll}
\tabletypesize{\scriptsize}
\tablecaption{2D imaging analysis of dual quasars and their hosts\label{tab:decomp}}
\tablehead{\colhead{ID}&\colhead{Quasar offsets\tablenotemark{a}}&\colhead{Separation\tablenotemark{b}}&\colhead{Quasar mag\tablenotemark{c}}&\colhead{Host mag\tablenotemark{c}}&\colhead{Host stellar mass}\\
&\colhead{(RA: \arcsec, Dec: \arcsec)}&\colhead{(\arcsec; kpc)}&\colhead{(HSC: $i$-band)}&\colhead{(HSC: $i$-band)}&\colhead{($\times10^{10}$ M$_{\odot}$)}}
\startdata
SDSS~J1214+0102&0.12, -0.26; -0.20, 1.9&2.2 (13.2)&20.22, 20.08&19.98, 20.40 &$7.6\pm2.8$, $3.8\pm1.3$\\
SDSS~J0847-0013&0.06, 0.06; 0.59, -0.73&1.0 (6.8)&19.15, 19.58&19.57&$8.5\pm4.6$\\
SDSS~J1416+0033 &0.38, -0.22; -0.22, -0.00&0.7 (3.9)&19.97, 19.94&18.80&$16.1\pm0.5$
\enddata
\tablenotetext{a}{These are spatial offsets relative to the position given in Table~\ref{tab:specsample}.}
\tablenotetext{b}{Projected separation given in arcsec and kiloparsecs.}
\tablenotetext{c}{Magnitudes resulting from the decomposition of the 2D HSC spatial profiles. Statistical errors are at the 0.01 mag level. Systematic uncertainties due to PSF mismatch have not been assessed.} 
\end{deluxetable*}

\section{Discussion: preliminary insights on the dual fraction of quasars}
\label{sec:dualfrac}

With a significant sample of dual quasar candidates drawn from a large parent population of quasars, we explore the frequency of dual quasar activity and compare with a simulated sample of dual quasars from Horizon-AGN, after roughly modeling the sample selection effects. Since followup spectroscopic confirmation is still in an early stage, we consider this estimate to have a substantial level of uncertainty. However, we report here on our current assessment of the dual quasar fraction since there are interesting constraints based on our candidates and observational results in the literature are also highly uncertain. Current observational estimates for the dual quasar rate are mainly limited to the low redshift ($z<0.1$) Universe at these close separations \citep[e.g.,][]{Liu2011,Koss2012}. 

\begin{figure}
\epsscale{2}
\plottwo{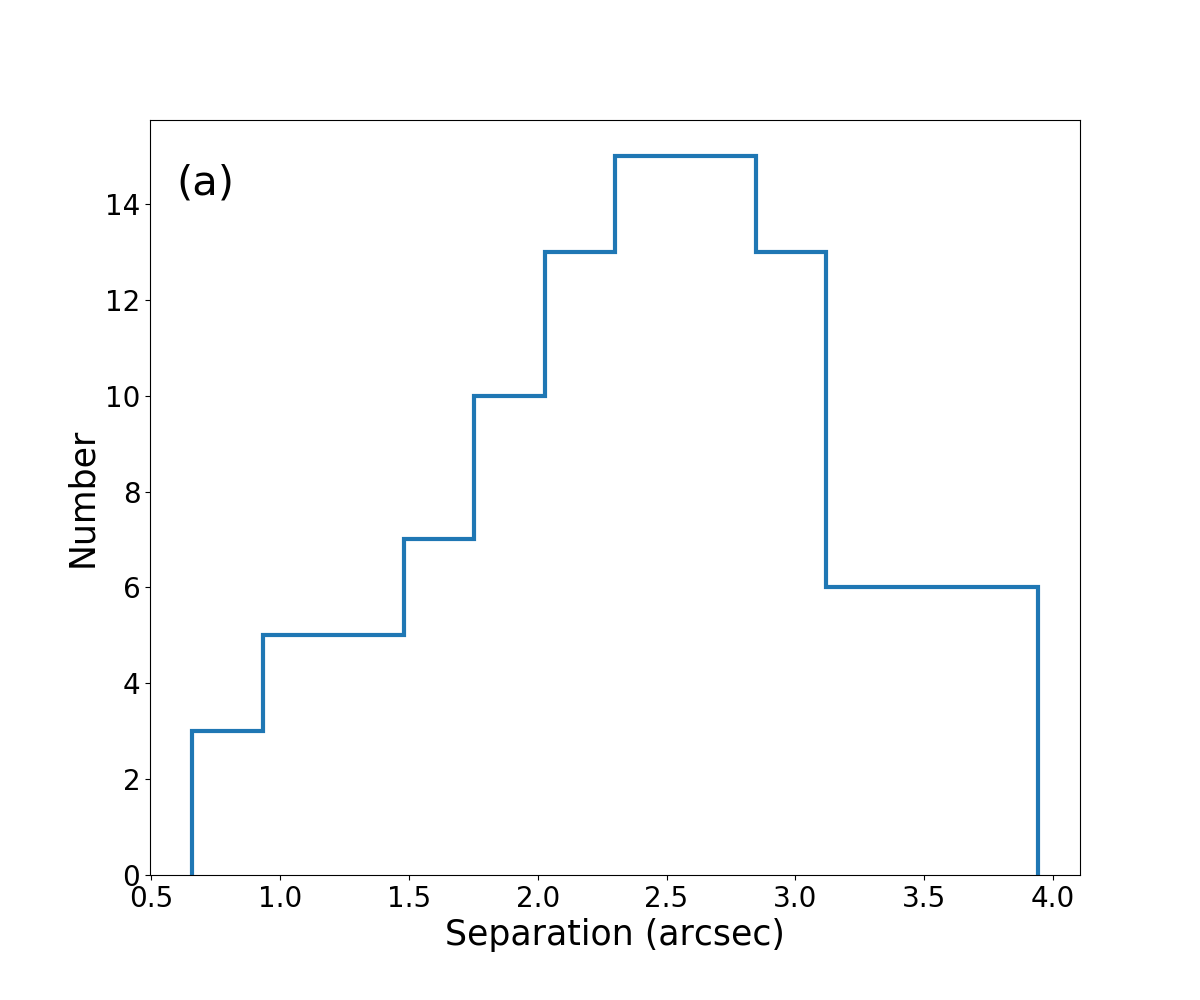}{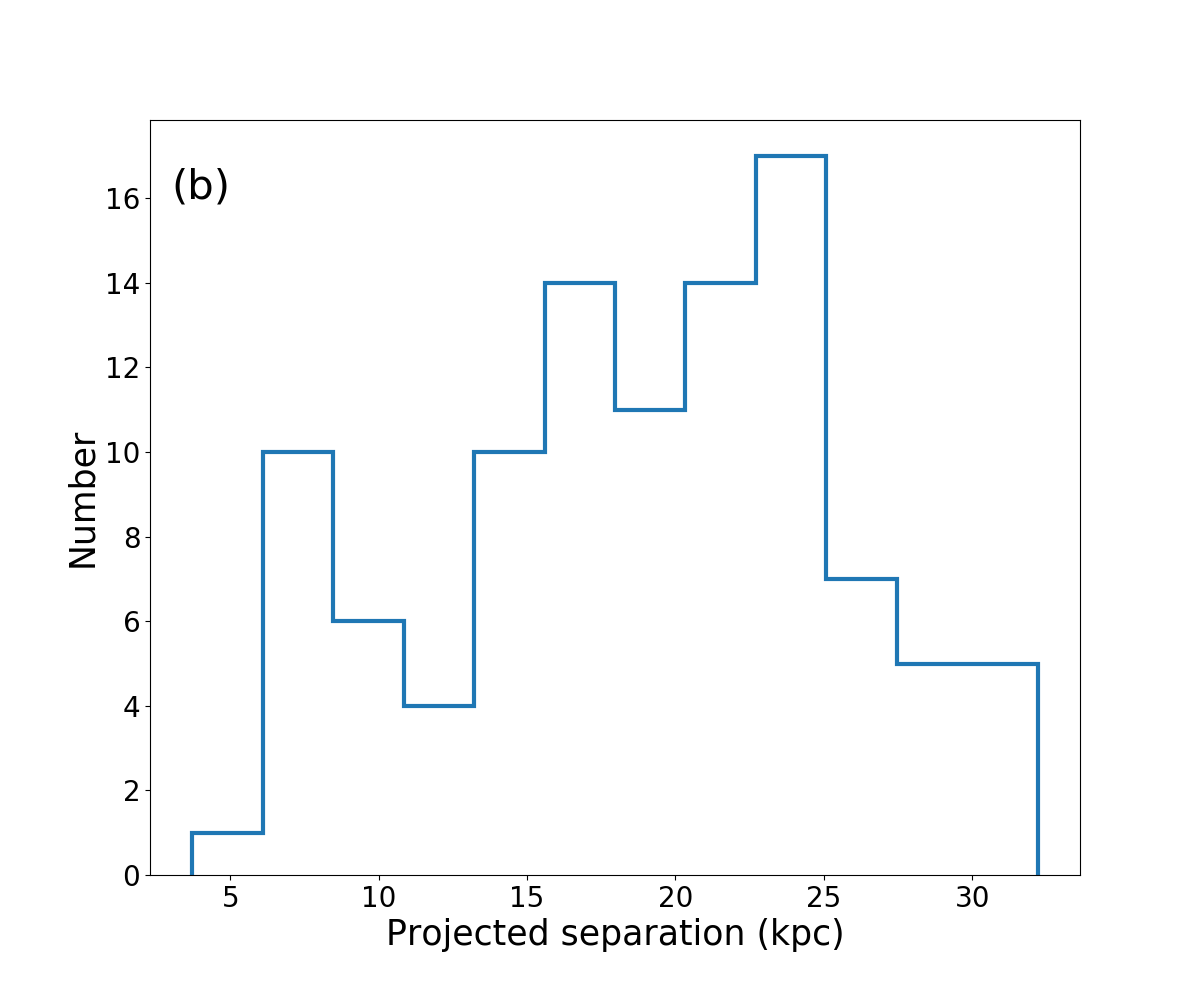}
\caption{(a) Angular separation in arcseconds of the 116 dual quasar candidates with companions having a blue color $g-r<1$. The projected physical distance (kpc) is shown in panel (b), assuming the components of each dual quasar candidate are at the same redshift. The distributions are based on observed quantities thus impacted by selection effects, particularly at smaller ($\lesssim$2$^{\prime\prime}$) separations, and dependent on the redshift distribution in the case of the physical separation.}
\label{fig:sample_sep}
\end{figure}

We measure a dual quasar fraction out to $z\sim3.5$ (i.e., across $\sim$12 billion years of cosmic time). The observed dual quasar fraction is the ratio of the number of objects meeting our criteria of dual quasar candidates (Section~\ref{sec:selection}) to the number of SDSS DR14 quasars with imaging from HSC SSP survey. Again, we consider a quasar to be part of a dual quasar system if it has a nearby companion at least 10\% as bright. For this exercise, we exclude cases where the companion has a red color ($g - r > 1$), solely to remove many stellar contaminants (Section~\ref{sec:stars}). The projected physical separation is restricted to be between 5--30 kpc. While our initial selection of dual quasar candidates included objects with separations between $0\farcs6$ and $4\farcs0$, we are likely incomplete below $1\farcs2$ (twice the median FWHM of the i-band PSF; Figure~\ref{fig:sample_sep}~top panel).  We thus limit ourselves to the 96 candidates (dual quasar systems) with separations between $1\farcs2$ and 4$^{\prime\prime}$.

There is a multiplicative correction factor that we apply to the observed dual quasar fraction to account for selection effects. We are likely to be missing dual quasar candidates at the smaller separations ($1\farcs2 - 2\farcs5$), even though we increased our lower limit from $0\farcs6$ to $1\farcs2$. As seen in Figure~\ref{fig:sample_sep} (top panel), there is a decline in the number of candidates with separations below $2\farcs5$. While we do not know the true distribution of the projected angular separation of dual quasars, we assume it to be flat over the range of separations considered here and peaking at $2\farcs5$. This amounts to a correction factor of 0.58. Hence, the observed dual quasar fraction is increased by a factor of 1.72.   

We then multiply the dual quasar fraction by the spectroscopic success rate based on initial results from our observing program with Keck and Gemini. With the requirement on the companion having $g-r<1$, there are three spectroscopically-confirmed dual quasar systems out of six candidates (Section~\ref{sec:spectroscopy}). The observed dual quasar fraction is then multiplied by a factor of 0.5. This fraction and its uncertainty is equally applied across all redshift bins. Therefore, there remains much uncertainty due to the limited sample of candidates with spectroscopy, especially at $z>1$. 

\begin{figure}
\epsscale{1.1}
\plotone{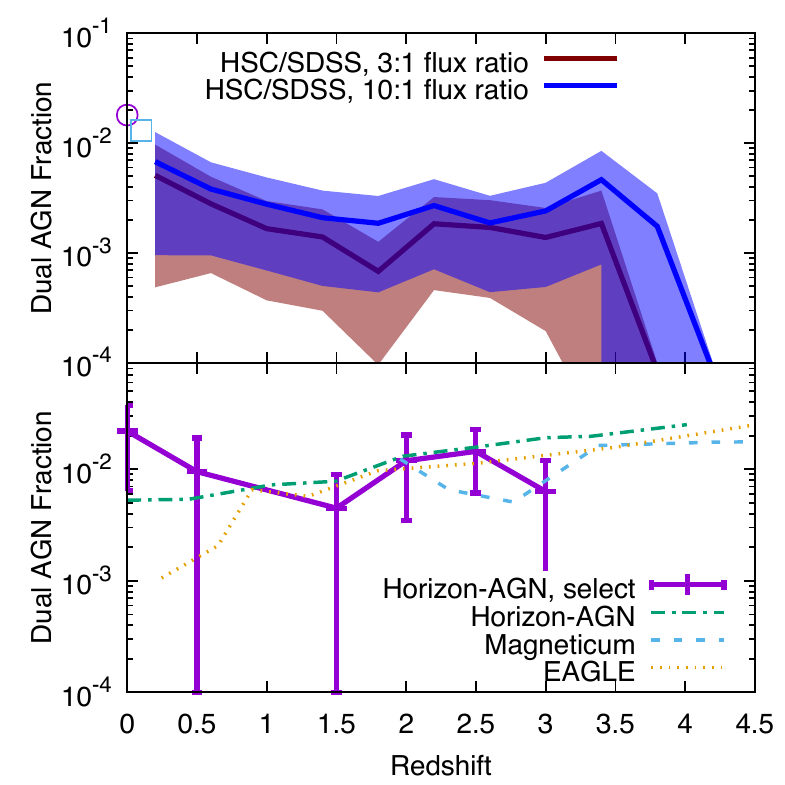}
\caption{Dual quasar fraction with projected separation between 5 and 30 kpc as a function of redshift. $Top$ Observational results from our Subaru/HSC program are displayed with the colored region indicating a 1$\sigma$ confidence interval based on Poisson statistics and the uncertainty on the spectroscopic success rate. The blue region is based on our sample having a flux ratio within 10:1 while the maroon region shows only those having a flux ratio within 3:1. For comparison, we show other published observational studies (thin light-blue square,  \citealt{Liu2011}; purple circle, \citealt{Koss2012}) that only exist for the local Universe ($z < 0.1$). $Bottom$ Results from cosmological hydrodynamic simulations based on real-space separations. Horizon-AGN \citep{Volonteri2016} is shown with a quasar selection matched to our observed sample (M. Volonteri, private comm.) thus labelled Horizon-AGN 'select'. For further comparison, we indicate the dual AGN fraction from the Magneticum \citep{Steinborn2016} and EAGLE \citep{Rosas-Guevara2019} simulations having a lower threshold in AGN luminosity that can lead to more pronounced differences in the dual AGN fraction, particularly at higher redshifts. The Magneticum, EAGLE and Horizon-AGN curves are based on the matched results shown in Figure 6 of \citet{DeRosa2019}.}
\label{fig:dualfraction}
\end{figure}

In Figure~\ref{fig:dualfraction}~$top~panel$, we plot the fraction of quasars that are dual with projected separation between 5--30 kpc. The statistical uncertainty of the dual quasar fraction is calculated in every redshift bin based on Poisson statistics. It is dominated by the total number of confirmed dual AGN and the number of dual AGN candidates per bin. Overall, we find a dual fraction of $0.26\pm0.18\%$, with no evidence of a dependence on redshift. If we limit ourselves to objects in which the two components have a flux ratio of 3:1 or less, the dual fraction is $0.16\pm0.12\%$, about a factor of 1.6 less, again showing no evidence for redshift evolution. Given the errors based on Poisson statistics and the considerable uncertainty in our spectroscopic success rate, particularly as a function of redshift, we cannot yet rule out the existence of some mild evolution. However, since we expect the spectroscopic success rate at higher redshift to drop given the greater likelihood of contamination due to projection effects and lensing, it seems unlikely that the fraction of dual quasars rises strongly with redshift, in contrast with expectations from simulations (see below).

Our dual quasar fraction rate is a factor of three higher than measured using dual quasar samples found from lens searches \citep{Kayo2012}. The difference may be because we relaxed the color criteria, allowing redder quasars (although still bluer than the stellar locus; Figure~\ref{fig:colors}), or because there are still foreground stars contaminating the sample. At $z < 1$, we find a dual quasar fraction considerably lower, by about a factor of four, than previously published observational results \citep{Liu2011,Koss2012}. This is also likely due to the varying selection of the quasar samples, highly impacted by luminosity and obscuration. Even so, a larger sample of spectroscopically confirmed dual quasars is needed at all redshifts. In particular, a better understanding of our selection function is needed since the number of pairs in our sample declines below $\sim2^{\prime\prime}$ (Figure~\ref{fig:sample_sep}), a separation larger than expected given the spatial resolution of the imaging in the $i$-band.

\subsection{Comparison with the Horizon-AGN and other simulations}

Cosmological hydrodynamic simulations now reach a large enough volume to allow a comparison to observational studies of luminous dual quasars. Here, we qualitatively compare the dual quasar fraction from observations and cosmological hydrodynamic simulations, particularly Horizon-AGN \citep{Volonteri2016}. The Horizon-AGN simulation \citep{Dubois2014} is based on the Adaptive Mesh Refinement Code RAMSES \citep{Teyssier2002} using a standard $\Lambda$CDM cosmology. The simulation box has a size of 100h$^{-1}$ Mpc$^{-1}$  on a side, 1024$^3$ dark matter particles, and an effective resolution of 1 kpc. 

Before proceeding, it is important to note that the SDSS quasar sample, used as our primary selection, is dominated by optically unobscured objects selected using a variety of color-based algorithms, and has a complex dependence on redshift and luminosity. This fact will likely contribute to differences between the observed and simulated samples. Not surprising, comparisons between simulations also require a careful matching of parameters as extensively discussed in \citet{DeRosa2019}.

To minimize such selection effects, we make an attempt to match the quasar properties of the Horizon-AGN simulation. We place a lower limit on the bolometric luminosity of the primary SDSS quasar at \hbox{log $L_\mathrm{primary} > 45.3$}. This limit is indicated in Figure~\ref{fig:sample}. We then require the companion to be within a factor of 10:1; this amounts to a luminosity limit of \hbox{log $L_\mathrm{secondary} > 44.3$}. A sample of simulated dual quasars (M.\ Volonteri, priv.\ comm) is constructed from Horizon-AGN meeting these luminosity thresholds and having real space separations of 5 -- 30 kpc. While it is beyond the scope of this study to fully assess the differences between projected and real-space separation, projections onto different axes within the simulation, evaluated as a simple check by assuming that the two quasars are at the same redshift, does not greatly change the results. We will more accurately model the selection in the simulations in future papers, particularly after we have obtained spectroscopic confirmation of a larger sample of dual quasar candidates. To avoid any confusion, we chose to plot here the observations and simulations separately in Figure~\ref{fig:dualfraction}.

As pointed out in \citet{DeRosa2019}, these comparisons should also be matched in black hole mass and the stellar mass of the underlying host galaxies. Therefore, we place lower limits on these quantities for dual quasars in the simulation based on measurements of these quantities from our three confirmed dual quasars. From model fits to the broad emission lines (Tang et al. in preparation), we report that their black hole masses fall between 10$^8$ and $10^9$ M$_{\odot}$. Therefore, we place a lower limit of 10$^8$ M$_{\odot}$ on both black holes in a dual system. Based on the stellar mass measurements of these dual quasars (Table~\ref{tab:decomp}), we place an additional constraint where  $M_{stellar}>10^{10}$ M$_{\odot}$ for both galaxies. With these cuts, we find 10 dual quasars in the Horizon-AGN simulation volume with $z \leq 3$. We label this matched simulated sample as Horizon-AGN 'select'.

As shown in Figure~\ref{fig:dualfraction}~$bottom~panel$, we find that the expected rates of dual quasar activity in the Horizon-AGN 'select' sample are about a factor of five higher than the observed sample of SDSS quasars with HSC. However, the errors on the simulated data are large, and thus the discrepancy is between 1 and 2$\sigma$,  particularly at $z < 2$. As mentioned above, the differences in normalization are expected since it is challenging to exactly match the selection between the full DR14 SDSS quasars catalog and quasars in the simulation. Putting aside the differences in rates, the simulation exhibits a lack of strong evolution similar to our observations.

We also provide the predictions of the dual quasar fraction from other cosmological hydrodynamic simulations \citep{Rosas-Guevara2019,Steinborn2016}. Unlike the Horizon-AGN sample, we haven't imposed the selection effects, thus these comparisons are more of an illustration of the impact of selection when comparing with lower luminosity AGNs that dominate the samples in these surveys. The EAGLE \citep{Rosas-Guevara2019} simulation has a higher rate than our observations by an order of magnitude (Figure~\ref{fig:dualfraction}), particularly at $z>1$, and exhibits an increase in the dual quasar fraction with redshift that is not seen in our observed dual quasar fraction or the Horizon-AGN 'select' simulation as mentioned above. Much of the discrepancy is likely due to the different luminosity thresholds: in the EAGLE simulation, AGN are defined to be objects with  hard ( 2--10 keV) X-ray luminosity above $10^{42}$ erg s$^{-1}$, a limit substantially lower than that of our SDSS sample. The Magneticum Pathfinder Simulations \citep{Steinborn2016} also show rates above our observed value at $z\sim2$ that is likely due to the different luminosity thresholds. To conclude, a comparison between observed and simulated samples with higher accuracy needs larger number statistics for these luminous dual quasars over a parameter space that includes redshift, projected separation, color, black hole mass, stellar mass, and Eddington accretion rate.

\section{Brief summary and concluding remarks}

We presented first results from an ongoing program to identify dual supermassive black holes, both being in a luminous quasar phase, using the Subaru HSC SSP. The exquisite quality of the deep and wide imaging of HSC in five optical bands facilitates the detection of dual quasars down to separations of a few kiloparsecs. We presented our selection of candidates from the known SDSS quasar population up to $z\sim4.5$ with consideration of the PSF and de-blended optical photometry with the latter important in providing color information needed to remove chance projections due to stars. With the use of Keck and Gemini in spectroscopy mode, we have confirmed three dual quasars, of which two were previously unknown. In the three cases presented here, the redshifts are so close to each other that spatially unresolved spectroscopy of the pair (e.g., from SDSS fiber spectroscopy) would not show a double-peaked [OIII]$\lambda$5007 line. In one case, we have found a pair of quasars, where one is unobscured while its close neighbor, just 3 kpc away, is moderately reddened. With the sample in hand, we have made a first attempt to quantify the dual fraction of quasars. The dual fraction of $0.26\pm0.18\%$ shows no evidence for a dependence on redshift. This may indicate that the triggering of simultaneous dual quasar activity is not any different at earlier epochs when the merger rates \citep[e.g., ][]{Lackner2014} and gas fractions \citep[e.g., ][]{Tacconi2020} of galaxies were significantly higher than today. Thus, the rates of dual quasar activity may simply be a result of the stochastic behavior of the instantaneous mass accretion rate onto the individual SMBHs in galaxy mergers \citep{Capelo2017,Goulding2018}. However, any firm conclusions require a larger spectroscopic sample and a better understanding of our selection function. This work is a demonstration of the improvements to be made in the study of dual quasars with larger samples with deep imaging to come from wide-area imaging surveys with HSC, LSST, Euclid and WFIRST. In addition to the luminous SDSS quasars, the optical imaging is deep enough to begin identifying dual quasars where both components are less luminous than those presented in this study.

\acknowledgments

We thank Marta Volonteri for providing the sample of dual quasars from the Horizon-AGN simulation and valuable guidance on the comparison between the observed and simulated data. Further input from Hugo Pfister was useful in this respect. We thank the anonymous referee for their valuable comments that improved the paper. We also recognize contributions from Thaisa Storchi Bergmann, Nadia Zakamska, and Jonelle Walsh in the reduction of the Gemini/NIFS data. JDS is supported by the JSPS KAKENHI Grant Number JP18H01521, and the World Premier International Research Center Initiative (WPI Initiative), MEXT, Japan. KGL is supported by the JSPS KAKENHI Grant Numbers JP18H05868 and JP19K14755. KI acknowledges support by the Spanish MICINN under grant PID2019-105510GB-C33. YM is supported by KAKENHI Grant Numbers 17H04831, 17KK0098, 19H00697, and 20H01953. R.A.R thanks partial financial support from Conselho Nacional de Desenvolvimento Cient\'ifico e Tecnol\'ogico (302280/2019-7) and Funda\c c\~ao de Amparo \`a pesquisa do Estado do Rio Grande do Sul (17/2551-0001144-9 and 16/2551-0000251-7).

\bibliographystyle{apj.bst}
\bibliography{references}

\end{document}